\documentclass[aps,physrev,preprint,groupedaddress]{revtex4-2}

\usepackage{dcolumn}
\usepackage{bm}

\usepackage{graphicx}
\usepackage{indentfirst}
\usepackage{braket}
\usepackage{float}
\usepackage{amsmath}
\usepackage{physics}
\usepackage{amssymb}
\usepackage{CJK}
\usepackage{esint}
\usepackage{color}
\usepackage[T1]{fontenc}
\usepackage{subfigure}
\usepackage{amsfonts}
\usepackage{footmisc}
\usepackage{scrextend}
\usepackage{multirow}
\usepackage[outdir=./]{epstopdf}
\usepackage{svg}
\usepackage{algorithm}
\usepackage{algpseudocode}
\usepackage{soul}
\usepackage{mathtools}
\usepackage[english]{babel}
\usepackage{url}
\usepackage{comment}
\usepackage{array}
\usepackage{subfiles}
\usepackage{tikz}
\usetikzlibrary{shapes,arrows}
\usepackage{mathrsfs}
\usepackage[mathscr]{eucal}
\usepackage[hyperfootnotes=false]{hyperref}
\usepackage{cleveref}
\usepackage{stmaryrd}

\definecolor{darkblue}{rgb}{0,0,0.5}
\hypersetup{
colorlinks=true,
linkcolor=black,
filecolor=blue,
citecolor=darkblue,  
urlcolor=black,
}

\usepackage{trimclip}

\makeatletter
\DeclareRobustCommand{\shortto}{%
  \mathrel{\mathpalette\short@to\relax}%
}

\newcommand{\short@to}[2]{%
  \mkern2mu
  \clipbox{{.5\width} 0 0 0}{$\m@th#1\vphantom{+}{\shortrightarrow}$}%
  }
\makeatother

\usepackage{pict2e,picture,graphicx}

\makeatletter
\DeclareRobustCommand{\Arrow}[1][]{%
\check@mathfonts
\if\relax\detokenize{#1}\relax
\settowidth{\dimen@}{$\m@th\rightarrow$}%
\else
\setlength{\dimen@}{#1}%
\fi
\sbox\z@{\usefont{U}{lasy}{m}{n}\symbol{41}}%
\begin{picture}(\dimen@,\ht\z@)
\roundcap
\put(\dimexpr\dimen@-.7\wd\z@,0){\usebox\z@}
\put(0,\fontdimen22\textfont2){\line(1,0){\dimen@}}
\end{picture}%
}
\makeatother

\newcommand{\Xin}[1]{{{\textcolor{orange}{#1}}}}

\def\be{\begin{equation}}
\def\ee{\end{equation}}
\def\ba{\begin{eqnarray}}
\def\ea{\end{eqnarray}}
\def\bal{\begin{equation}\begin{aligned}}
\def\eal{\end{aligned}\end{equation}}

\def\bp{\begin{pmatrix}}
\def\ep{\end{pmatrix}}

\usepackage{bm}

\newcommand{\calA}{{\cal A}}

\newcommand{\calT}{{\cal T}}

\newcommand{\calJ}{{\cal J}}

\newcommand{\calR}{{\cal R}}

\usepackage[normalem]{ulem}

\begin{document}


\title{\textbf{Mode Selection in Quantum Nonlinear Optics Using Optical Resonators} 
}%

\author{Xin Chen}%
 \email{chenxin@qzc.edu.cn}
\affiliation{%
 The College of Electrical and Information Engineering, Quzhou University, Quzhou 324000, China.
}%

\date{\today}




\begin{abstract}
Nonlinear optics underpins quantum photonics by enabling the generation and control of quantum states of light. We present new applications of optical resonators as mode selectors in nonlinear processes. First, we show that cavity-enhanced spontaneous parametric down-conversion can generate spectrally uncorrelated photon pairs with improved decorrelation and wavelength flexibility. Second, we demonstrate that a cavity-assisted sum-frequency generation process realizes a quantum pulse gate with high-resolution temporal-mode selectivity and precise spectral control. Our theoretical framework provides a general methodology for analyzing cavity-enhanced nonlinear processes and highlights the versatility of optical resonators as powerful tools for engineering quantum light.
\end{abstract}


\maketitle

Nonlinear optics underpins the development of quantum photonics \cite{Chang2014} and quantum information science \cite{Nielsen2000,Clark2015}, providing the essential mechanisms for generating, manipulating, and detecting quantum states of light. A prime example is spontaneous parametric down-conversion (SPDC) \cite{ou2007multi,PhysRevA.50.5122}, widely used to produce entangled photon pairs that serve as key resources for quantum communication \cite{Takeda2013,PhysRevLett.117.240503,gisin2007quantum,wehner2018quantum,wilde2013quantum,kimble2008quantum}, cryptography \cite{Ekert_1991}, computing \cite{Su2013,Shor}, and quantum-enhanced sensing \cite{toth2014quantum,zhang2021dqs,Brady2023,giovannetti2006,giovannetti2011advances}. Other nonlinear processes such as sum-frequency generation (SFG) and its inverse enable quantum frequency conversion \cite{Allgaier2017,PhysRevLett.109.147404}, a crucial function for interfacing disparate quantum systems including memories, processors, and telecom networks.

Optical resonators (cavities) are ubiquitous and versatile elements in photonics. They provide spatial and spectral mode selection—acting as narrowband filters—while enhancing the optical field within a confined volume. This enhancement is particularly important for driving efficient nonlinear interactions, placing resonators at the core of devices such as lasers, optical parametric oscillators, and other three- and four-wave mixing systems \cite{Bayvel1989,Levy2009,PhysRevApplied.17.034012,PhysRevLett.123.193603}.

Here we investigate new roles of optical cavities as mode selectors in nonlinear processes, focusing on two cases. The first involves cavity-enhanced SPDC (CSPDC) to generate spectrally uncorrelated photon pairs, a key requirement for producing pure heralded single photons \cite{PhysRevLett.100.133601,Li2020,single-photon-ren}—fundamental resources for quantum optics and quantum information processing. This configuration, resembling a broadband-pumped optical parametric oscillator, allows the cavity to act simultaneously as a spectral filter and field enhancer, offering improved spectral decorrelation and greater flexibility in wavelength and temporal/spectral shaping compared with existing methods. The second case concerns cavity-enhanced SFG (CSFG) for realizing a quantum pulse gate (QPG) \cite{Eckstein:11,PhysRevA.90.030302,Reddy:14,Reddy:18}, which acts as a temporal-mode–selective beam splitter: extracting one target temporal mode (TM) \cite{PhysRevX.5.041017,Raymer_2020} from a multimode input and converting it into another frequency while leaving all others unaffected. Extension to a multi-output QPG (MQPG) \cite{PRXQuantum.4.020306,PRXQuantum.5.040329} enables complete decoding in high-dimensional quantum communication. Compared with conventional approaches, the cavity-based configuration provides finer control over the target mode profile and achieves high-resolution frequency selectivity, overcoming the usual limitations of spectral engineering and time-ordering distortions in the high-conversion regime \cite{Reddy:13}.

Finally, while resonators are conventionally treated via Langevin equations in the Heisenberg picture with continuous frequency modes, we develop a complementary approach that employs either the interaction picture—providing an intuitive view of spectral correlations—or the Heisenberg picture—yielding accurate solutions in the strong-interaction regime. Fields can be expanded in continuous or discrete frequency domains, a flexibility that enables addressing different situations. This establishes a general foundation for resonator-based nonlinear processes.

{\em Spectrally uncorrelated photon pair via cavity-enhanced SPDC.---}
We begin by outlining the model of CSPDC, as illustrated in Fig.~\ref{fig:spdc}. For simplicity, all fields are assumed to occupy a single spatial mode (not explicitly shown in the figure for clarity), achievable, for example, using a waveguide. A strong, undepleted pump pulse, proportional to
$
\beta(t) = (1/\sqrt{2\pi}) \int \beta(\omega_p) {\rm e}^{-i\omega_p t} \, d\omega_p,
$
is incident on a nonlinear medium such as a crystal or a nonlinear fiber. Through the second-order interaction, a portion of the pump photons is down-converted into signal–idler photon pairs. The signal and idler fields are confined within independent optical cavities with decay rates $\gamma_j$, which enhance the resonant modes while suppressing off-resonant frequencies, where $j \in \{s,i\}$ denotes signal ($s$) and idler ($i$). The corresponding intracavity mode operators are $\hat{b}_j$ with central frequencies $\omega_{j,c}$. The external field modes coupled to the cavities are defined as
$\hat{a}_j=\sqrt{1/2\pi}\int d\omega_j\,\hat{a}_{j}(\omega_j)$. By appropriately choosing the cavity decay rates and free spectral ranges, only a single resonance of each cavity contributes significantly. The polarizations and propagation directions of the interacting fields are arranged to satisfy both energy conservation and phase-matching conditions.

\begin{figure}[h]
    \centering
 \includegraphics[width=0.6\linewidth]{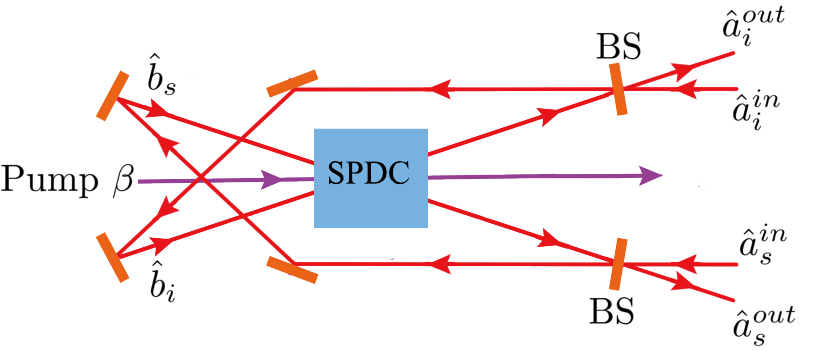}
    \caption{Schematic illustration of the CSPDC. `BS': Beamsplitter. }
    \label{fig:spdc}
\end{figure}

In the rotating frame defined by $\hat{o}_{j}(\omega_j)\rightarrow\hat{o}_{j}(\omega_j) {\rm e}^{-i\omega_{j,c} t}$, $\beta(t)\rightarrow\beta(t){\rm e}^{-i\omega_{p,c} t}$  with $o\in\{a,b\}$ and the carrier frequencies satisfying $\omega_{p,c} = \omega_{s,c} + \omega_{i,c}$, the Hamiltonian reads 
\bal
\hat{H}=\hat{H}_0+\hat{H}_1,
\label{Hamitonianallro1}
\eal
where $\hat{H}_0/\hbar=\sum_{j}[\int d\omega_j\, \omega_j\,\hat{a}^{\dagger}_{j}(\omega_j)\hat{a}_{j}(\omega_j)+i\sqrt{\gamma_j} (\hat{a}_j^{\dagger}\hat{b}_j-\hat{b}_j^{\dagger}\hat{a}_j)]$ and $\hat{H}_1/\hbar=-i \eta [\hat{b}_s^{\dagger}\hat{b}_i^{\dagger}\beta(t)-\hat{b}_s\hat{b}_i\beta^{*}(t)]$. We assume the fields are one-dimensional and quasi-monochromatic \cite{ou2017quantum}. The spatial integrals in Hamiltonian have already been carried out. In $\hat{H}_1$, the resulting phase-matching factor $C=\int_0^L {\rm exp}(i\Delta k z)dz$ is treated as constant by tailoring the cavity decay rates such that only the near-resonant frequencies satisfying $\Delta k L \ll 1$ are enhanced, while off-resonant frequencies are suppressed. Here $\Delta k = k_p - k_s - k_i$, with $k_p$, $k_s$, and $k_i$ the wavevectors of the pump, signal, and idler fields, and  $L$ is the interaction length. For conciseness, Schr\"{o}dinger-picture subscripts $S$ are omitted. The broadband pump ($\Omega_p \gg \max(\gamma_s,\gamma_i)$) is modeled with a scaled spectral profile $\beta(\omega_p)=\Pi(\omega_p/\Omega_p)$, where $\Pi$ denotes the rectangular function. Its overall scale and the spatial overlap integral $C$ are absorbed into the effective nonlinear coupling $\eta$, which is also proportional to the second-order nonlinear susceptibility $\chi^{(2)}$. Without loss of generality, we take $\eta$ to be positive and real. In the weak-conversion regime—relevant for heralded single-photon generation, where multipair contributions must be suppressed—$\hat{H}_1$ is treated perturbatively in the interaction picture. The intracavity mode operators in the interaction picture then follow as \cite{supp}
\be
\hat{b}_{j,I}(\omega_j)=\sqrt{2\pi}\xi^-_{j}(\omega_j)\hat{a}^{in}_{j,I}(\omega_j),
\label{cavityr}
\ee
with the input–output relation 
\be
\hat{a}^{out}_{j,I}(\omega_j)=\frac{\xi^{-}_{j}(\omega_j)}{\xi^{+}_{j}(\omega_j)}\hat{a}^{in}_{j,I}(\omega_j),
\label{inoutr2}
\ee
where $\xi^{\mp}_{j}(\omega_j)=\sqrt{\gamma_j/2\pi}/(i\omega_j\mp\gamma_j/2)$. The input and output operators are defined in the interaction and Heisenberg pictures as 
\(\hat{a}_{j,q}^{in(out)}(\omega_j) = \lim_{t \to \mp \infty} \hat{a}_{j,q}(\omega_j,t) e^{i \omega_j t}\) 
(\(q \in \{I,H\}\)), and in the Schrödinger picture as 
\(\hat{a}_{j,S}^{in}(\omega_j) = \lim_{t \to -\infty} \hat{a}_{j,S}(\omega_j) e^{i \omega_j t}\), corresponding to the frequency-domain fields incident on and emitted from the cavity. 
By construction, these input operators coincide, 
\(\hat{a}_{j,S}^{in}(\omega_j) = \hat{a}_{j,I}^{in}(\omega_j) = \hat{a}_{j,H}^{in}(\omega_j)\).
 The time-evolution operator in the interaction picture $U_{1,I}(-\infty,\infty)$ is then approximated as
\bal
U_{1,I}(\infty,-\infty)&\approx {\rm exp}\bigg(-\frac{i}{\hbar}\int_{-\infty}^{\infty}\hat{H}_{1,I} dt\bigg)\\&\approx{\rm exp}(\sqrt{2\pi}\eta  \hat{A}_{s,I}^{in}\hat{A}^{in}_{i,I}-h.c.)
\label{Uint}
\eal
where $\hat{H}_{1,I}$ is the interaction-picture representation of the perturbation Hamiltonian $\hat{H}_{1}$, the TMs are defined as $\hat{A}_{j,q'}^{in(out)}=\int \xi^{\mp}_{j}(\omega_j) \hat{a}_{j,q'}^{in(out)}(\omega_j)d\omega_j$ with $q'\in \{S,I,H\}$ (note that $\hat{A}_{j,S}^{out}$ is not defined), and ``$h.c.$'' stands for hermitian conjugate. Since all $\hat{a}_{j,q'}^{in}$ are identical, Eq. (\ref{inoutr2}) implies that $\hat{A}_{j,I}^{out}$ equals all three corresponding input modes $\hat{A}_{j,q'}^{in}$.

\begin{figure}[t]
    \centering
 \includegraphics[width=0.6\linewidth]{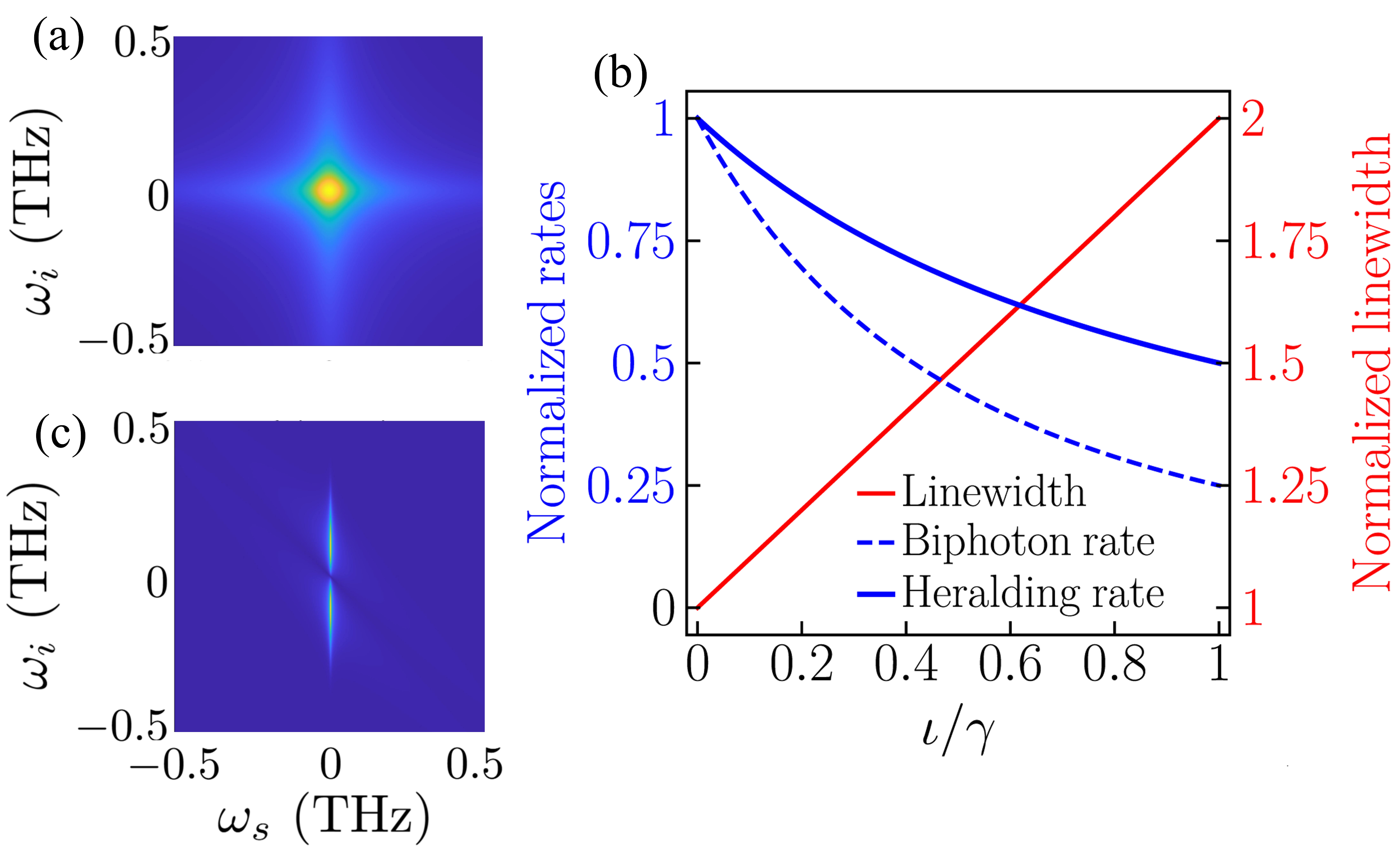}
    \caption{(a) Amplitude of the JSF for Type-II SPDC in BBO pumped by a broadband field centered at 263 nm, with both signal and idler resonant at 525 nm ($\gamma_j=0.1$ THz); idler mode purity $p=0.9996$. (b) Normalized linewidth, biphoton generation rate, and heralding rate versus internal loss $\iota/\gamma$ for equal losses in signal and idler modes; all quantities are normalized to their respective lossless values. (c) Amplitude of the JSF for the single-resonator configuration with pump in the second Hermite–Gaussian mode ($\gamma_s=0.004$ THz). Colormaps in (a) and (c) are scaled individually.}
    \label{fig:2}
\end{figure}

The corresponding Heisenberg-picture mode operator evolves as \cite{supp}
\bal
\hat{A}_{j,H}^{out}&=U_{1,I}^\dagger(\infty,-\infty)\hat{A}_{j,I}^{out}U_{1,I}(\infty,-\infty)\\&={\rm cosh}(\sqrt{2\pi}\eta)\hat{A}^{in}_{j,H}-{\rm sinh}(\sqrt{2\pi}\eta)\hat{A}^{in\dagger}_{\Bar{j},H},
\label{inoutr}
\eal
where $\bar{j}$ denotes the complementary mode ($\bar{s}=i$ and $\bar{i}=s$). From Eq. (\ref{inoutr}), the Schr\"{o}dinger-picture evolution operator is $U_S(\infty,-\infty)=\exp(\sqrt{2\pi}\eta  \hat{A}^{in}_{s,S}\hat{A}^{in}_{i,S}-h.c.)$. The corresponding output state is $\ket{\Phi}\approx\sqrt{1-2\pi\eta^2}\ket{vac}-\sqrt{2\pi}\eta\ket{1_s,1_i}$, with a biphoton probability $2\pi\eta^2$. Fig.~\ref{fig:2}(a) shows the amplitude of the joint spectral function (JSF) \cite{xinchen_pra2020} of a CSPDC process, where the degree of signal–idler decorrelation is quantified indirectly by the idler-mode purity $p=0.9996$, defined as $p = \mathrm{Tr}(\hat{\rho}_i^2)$ \cite{PhysRevLett.100.133601}, with $\hat{\rho}_i$ the idler density operator. For experimental feasibility, the cavity decay rate $\gamma_j$ must satisfy $\Omega_{SPDC}/F_j < \gamma_j < \Omega_{PM}$, where $\Omega_{PM}$ (typically 0.1-1~THz \cite{Chen:25}) is the bandwidth over which the quasi-phase-matching condition $\Delta k L \ll 1$ holds, ensuring the spatial integral $C$ is approximately constant; $\Omega_{SPDC}$ (typically 1-10~THz \cite{ou2007multi,Chen:25}) is the intrinsic SPDC bandwidth without cavity enhancement; and $F_j = \Omega_{FSR,j}/\gamma_j$ denotes the cavity finesse. This condition ensures that the free spectral range $\Omega_{FSR,j}$ exceeds $\Omega_{SPDC}$, so that only a single cavity resonance is enhanced. This scheme offers greater wavelength flexibility than dispersion-engineered approaches, as the cavity resonances—easily tunable over a broad range—set the uncorrelated wavelengths.

We now include internal cavity losses (e.g., scattering and absorption), modeled as coupling to an auxiliary input field $\hat{d}_j^{in}$ with coupling rate $\iota_j<\gamma_j$. 
The frequency correlations of the outputs are captured by the biphoton wavefunction (JSF) in the Heisenberg picture \cite{supp}:
\bal
\calJ(\omega_s,\omega_i)&=\bra{0}\hat{a}^{out}_{s,H}(\omega_s)\hat{a}^{out}_{i,H}(\omega_i)\ket{0}\\&\approx-\sqrt{2\pi}\eta\tilde{\xi}^-_s(\omega_s)\tilde{\xi}^-_i(\omega_i),
\eal
with 
$\tilde{\xi}^-_j(\omega_j)=\sqrt{\gamma_j/2\pi}/{(i\omega_j-\gamma_j/2-\iota_j/2)}$. The corresponding biphoton probability is  $p_{si}=\int |\calJ(\omega_s,\omega_i)|^2d\omega_sd\omega_i=2\pi\eta^2\gamma_s\gamma_i/(\gamma_s+\iota_s)(\gamma_i+\iota_i)$. Internal losses thus lead to a broadening of the signal and idler linewidths, 
and to a reduction of the biphoton generation probability.
However, they do not introduce additional spectral correlations, provided that the quasi-phase-matching condition 
$\Delta k L \ll 1$ remains satisfied over the entire broadened bandwidth. In addition, losses generate unpaired single photons, reducing the heralding efficiency to $\calR_j=p_{si}/p_j={\gamma_{\bar{j}}}/(\gamma_{\bar{j}}+\iota_{\bar{j}})$, where $p_j=\int \bra{0}\hat{a}^{out\dagger}_{j,H}(\omega_j)\hat{a}^{out}_{j,H}(\omega_j)\ket{0}d\omega_sd\omega_i$ is the total probability of detecting a photon in channel $j$, regardless of whether its partner in $\bar{j}$ is present. To illustrate how internal losses affect system performance, we assume equal normalized losses for the signal and idler modes, 
i.e., $\iota_j/\gamma_j = \iota/\gamma$.  
Fig. \ref{fig:2}(b) shows the normalized linewidth, biphoton generation rate, 
and heralding rate versus $\iota/\gamma$.

A variant of the CSPDC scheme is to apply resonant enhancement only to the signal field. The corresponding Schr\"{o}dinger-picture time-evolution operator is
$
U_{S}(\infty,-\infty)\approx{\rm exp}(\eta  \hat{A}^{in}_{s,S}\hat{A}^{in}_{i,S}-h.c.)
$, with TMs $\hat{A}^{in}_{s,S}=\int \xi^-_{s}(\omega_s) \hat{a}^{in}_{s,S}(\omega_s)d\omega_s$ and $\hat{A}^{in}_{i,S}=\int \beta(\omega_i) \hat{a}^{in}_{i,S}(\omega_i)d\omega_i$ \cite{supp}. By tailoring the pump $\beta(t)$, the idler output mode directly inherits the desired temporal profile, enabling flexible spectral shaping and benefiting heralded single-photon applications. Fig.~\ref{fig:2}(c) illustrates the amplitude of a representative JSF for this configuration.

\begin{figure}[t]
    \centering
    \includegraphics[width=0.6\linewidth]{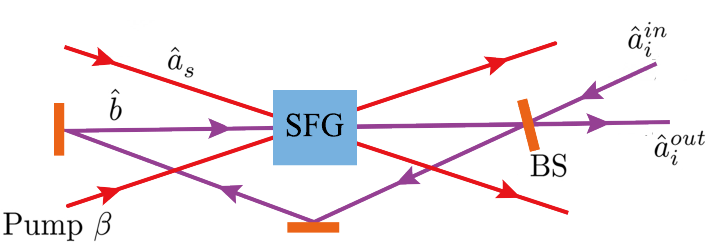}
    \caption{Schematic illustration of the CSFG. `BS': Beamsplitter. }
    \label{fig:qpg}
\end{figure}

{\em Quantum pulse gating via cavity-enhanced SFG.---}
We consider a model of the CSFG process (Fig.~\ref{fig:qpg}) with all fields in a single spatial mode. A strong, undepleted pump pulse with normalized temporal profile $\beta(t)$, defined such that 
$\int_{-T/2}^{T/2}|\beta(t)|^2 dt = 1$, drives a $\chi^{(2)}$ nonlinear medium. In the SFG interaction, a pump photon and a signal photon ($\hat{a}_s$, centered at $\omega_{s,c}$) are up-converted into a single idler photon at their sum frequency. The idler channel is resonantly enhanced by a cavity at frequency $\omega_{i,c}$, with intracavity mode $\hat{b}$ coupled to the external field $\hat{a}_i$. Assuming the pump and signal bandwidths are narrower than the free spectral range $\Omega_{FSR}$, only a single cavity resonance contributes. By appropriately choosing carrier frequencies and polarizations, energy conservation and phase matching are satisfied. As in the CSPDC case, we aim to remove frequency correlations between the signal and idler fields. While this is not possible in the continuous-frequency representation (infinite time window) in the high-conversion regime, it can be achieved with a finite-duration field of length $T$ (discrete-frequency representation), which is appropriate for practical applications where only a finite number of TMs within the relevant bandwidth participate.

In the rotating frame defined by $\hat{a}_{j}(\omega_{n_j})\rightarrow\hat{a}_{j}(\omega_{n_j}){\rm e}^{-i\omega_{j,c} t}$, $\hat{b}\rightarrow\hat{b}{\rm e}^{-i\omega_{i,c} t}$, $\beta(t)\rightarrow\beta(t){\rm e}^{-i\omega_{p,c} t}$ with $\omega_{s,c}+\omega_{p,c}=\omega_{i,c}$, the system Hamiltonian reads
\bal
\hat{H}=\hat{H}_0+\hat{H}_1,
\label{Hamitonianallro2}
\eal
where
$
\hat{H}_0/\hbar=\sum_{j,n_j} \omega_{n_j}\hat{a}^{\dagger}_{j}(\omega_{n_j})\hat{a}_{j}(\omega_{n_j})+i\sqrt{\gamma} (\hat{a}^{\dagger}_i\hat{b}-\hat{b}^{\dagger}\hat{a}_i)
$
and
$
\hat{H}_1/\hbar=-i \eta [\hat{a}_s\hat{b}^{\dagger}\beta(t)-\hat{a}_s^{\dagger}\hat{b}\beta^{*}(t)]
$
with $\hat{a}_j=\sqrt{1/T}\sum_{n_j} \hat{a}_{j}(\omega_{n_j})$, $\omega_{n_j}=n_j\Delta\omega$ and $\Delta\omega=2\pi/T$. The spatial integral is taken as approximately constant by suitably choosing the pump and signal bandwidths. Here, $\eta>0$ is the effective nonlinear coupling strength and $\gamma$ is the cavity decay rate. The system is analyzed in both the low- and high-conversion regimes.

For weak interaction, we treat $\hat{H}_1$ as a perturbation in the interaction picture. The interaction-picture evolution operator reads
$
U_{1,I}(T/2,-T/2)\approx {\rm exp}\big(\sum_{n_i=-\infty}^{\infty}\rho_{n_i} \hat{a}^{in}_{i,I}(\omega_{n_i}) \hat{A}^{in\dagger}_{s,I}(n_i)-h.c.\big),
\label{teo2}
$
where $\rho_{n_i}=\eta\sqrt{\gamma/T}/(i\omega_{n_i}-\gamma/2)$ \cite{supp}. The TMs are defined as $\hat{A}^{in}_{s,q}(n_i)=\sum_{n_s}\beta(\omega_{n_i-n_s})\hat{a}^{in}_{s,q}(\omega_{n_s})$ with $\hat{a}^{in(out)}_{j,q}(\omega_{n_j})=\hat{a}_{j,q}(\omega_{n_j},\mp T/2){\rm e}^{\mp i\omega_{n_j} T/2}$. Hence, the input-output relation for the idler channel is  
\bal
\hat{a}^{out}_{i,H}(\omega_{n_i})={\rm cos}(|\rho_{n_i}|)\hat{a}^{in}_{i,H}(\omega_{n_i})+{\rm sin}(|\rho_{n_i}|)\hat{A}^{in}_{s,H}(n_i).
\label{inout3}
\eal
The separability 
${\rm sin}^2|\rho_0|/\sum_{n_i}{\rm sin}^2|\rho_{n_i}|$ \cite{Reddy:13} quantifies the fraction of the converted target mode $\hat{A}^{in}_{s,H}(0)$ relative to all converted modes $\{\hat{A}^{in}_{s,H}(n_i)\}$, approaching unity in the limit $\gamma T \rightarrow 0$ ($\gamma / \Delta\omega \rightarrow 0$). By definition, the target TM $\hat{A}^{\rm in}_{s,H}(0)$ is fully shaped by the pump spectrum, highlighting that the pump’s spectral profile dictates the mode that is converted. A limitation of this perturbative, interaction-picture treatment is that the conversion efficiency (CE) \cite{Reddy:13} of the target TM, given by ${\rm sin}^2|\rho_0|$, remains low owing to the assumption of weak interaction; this does not reflect practical regimes, where higher efficiencies are generally required.
\begin{figure}[t]
    \centering
    \includegraphics[width=0.6\linewidth]{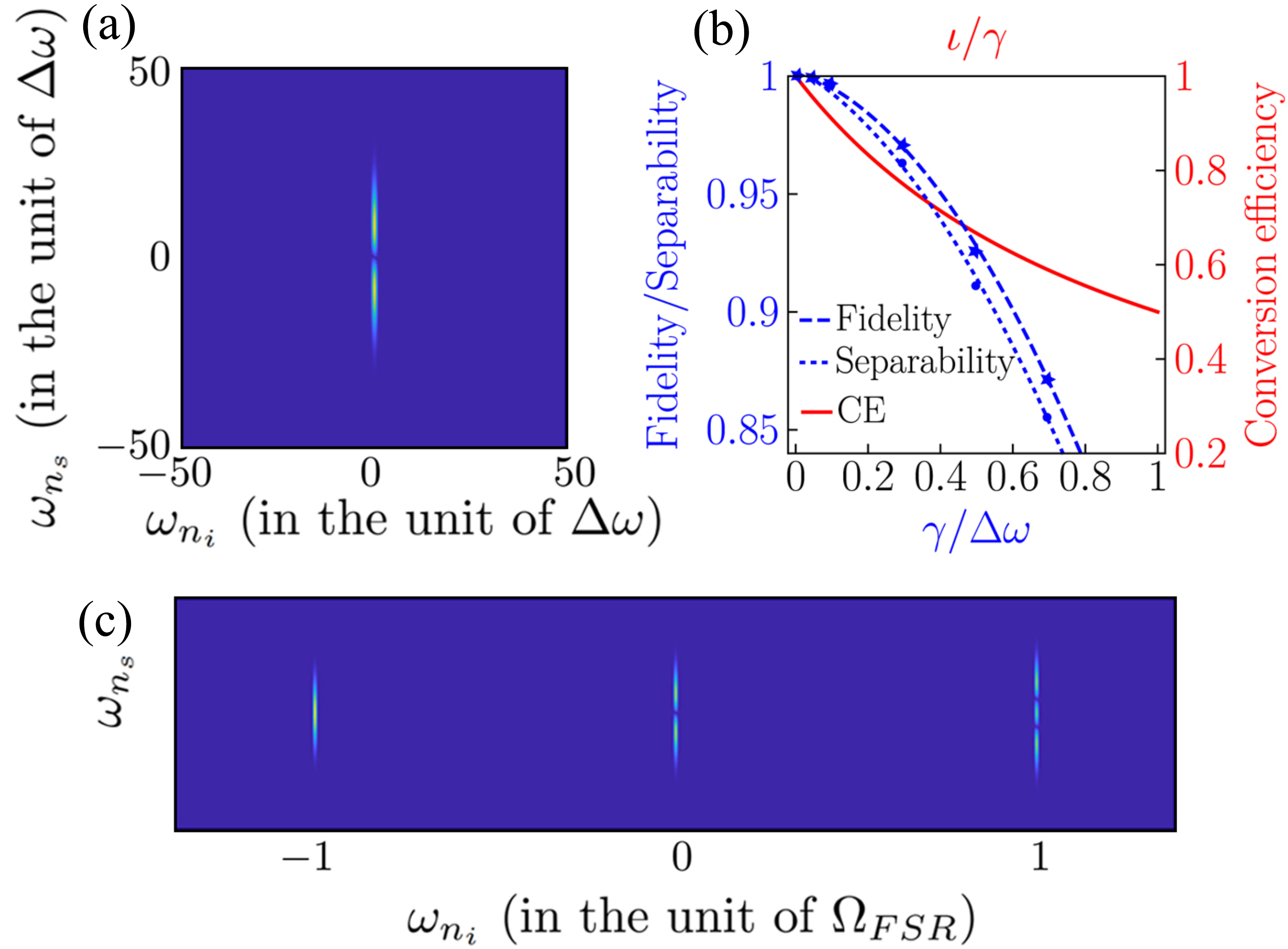}
    \caption{(a) Amplitude of the frequency-domain transfer function $G_s(\omega_{n_i},\omega_{n_s})$ for CSFG. Parameters: $\gamma/\Delta\omega=\eta^{2}/2\pi=10^{-2}$. For numerical tractability, 100 frequency modes are used, with the pump spectrum in the second Hermite–Gaussian mode. (b) Separability of the target TM and its fidelity relative to the pump field $\beta(t)$ as functions of $\gamma/\Delta\omega$, with all other parameters fixed as in (a) and $\gamma/\Delta\omega$ varied under the constraint $\eta =\sqrt{\gamma T}$. Also shown is the CE of the target TM, as a function of the internal loss $\iota/\gamma$ in the limit $\eta=\sqrt{(\gamma+\iota)T}\to0$. (c) Amplitude of the three-output transfer function of a three-peak MQPG, with pump peaks shaped as the first three Hermite–Gaussian modes.}
    \label{fig:4}
\end{figure}

To explore the high-conversion regime, it is necessary to work in the Heisenberg picture, where the issue of Hamiltonian time ordering can be properly addressed. From Eq.~(\ref{Hamitonianallro2}), the Heisenberg–Langevin equation for the cavity mode $\hat{b}$ is
\be
\dot{\hat{b}}(t)=-\frac{\gamma}{2}\hat{b}(t)-\eta\hat{a}_s^{in}(t)\beta(t)-\frac{\eta^2}{2}\hat{b}(t)|\beta(t)|^2-\sqrt{\gamma}\hat{a}_i^{in}(t),
\label{ceq3}
\ee
with periodic boundary condition $\hat{b}(T/2)=\hat{b}(-T/2)$. Input/output operators are defined in the time domain as $\hat{a}_j^{in(out)}(t)=(1/\sqrt{T})\sum_{n_j} \hat{a}^{in(out)}_j(\omega_{n_j}) {\rm e}^{-i\omega_{n_j} t}$. For clarity, we omit the subscript $H$ denoting Heisenberg operators. 

In Ref.~\cite{Chen:25}, we examined a special case where the pump spectral amplitudes $\beta(\omega_{-n_p})$ are modeled as a large set of random numbers drawn from an identical complex Gaussian distribution. In the limit $M\to\infty$, this yields $|\beta(t)|^2=1/T$, since ${\rm lim}_{M\to\infty}\sum_{k_p=1}^M\beta(\omega_{k_p-n_p})\beta^*(\omega_{k_p-m_p})=\delta_{m_p,n_p}$. Solving Eq.~(\ref{ceq3}) in the frequency domain and applying the input–output relation
$\hat{a}_i^{out}(t)-\hat{a}_i^{in}(t)=\sqrt{\gamma}\hat{b}(t)$,
we obtain the idler output field
$\hat{a}^{out}_{i}(\omega_{n_i})=\mu_{n_i}\hat{A}_s^{in}(n_i)+\nu_{n_i}\hat{a}^{in}_{i}(\omega_{n_i})$,
with coefficients $\mu_{n_i}=\eta\sqrt{\gamma/T}/(i\omega_{n_i}-\gamma/2-\eta^2/2T)$ and $\nu_{n_i}=(i\omega_{n_i}+\gamma/2-\eta^2/2T)/(i\omega_{n_i}-\gamma/2-\eta^2/2T)$, which satisfy $|\mu_{n_i}|^2+|\nu_{n_i}|^2=1$ to preserve unitarity. The separability
and target-TM CE are given by $|\mu_{0}|^2/\sum_{n_i}|\mu_{n_i}|^2$ and $|\mu_{0}|^2$, respectively. In the limit $\eta=\sqrt{\gamma T}\to 0$ ($\gamma/\Delta\omega\rightarrow 0$), the output simplifies to
\be
\hat{a}^{out}_{i}(\omega_{n_i})=
\begin{cases}
    -\hat{A}_s^{in}(0) & \text{if} \ n_j=0\\
    \hat{a}^{in}_{i}(\omega_{n_j}) & \text{if} \ n_j\neq0.
\end{cases}
\label{fullc}
\ee
Both the separability and CE approach unity. In this regime, the target mode $\hat{A}^{in}_s(0)$ is fully converted, while all orthogonal input modes are transmitted unchanged. This limiting case serves as a useful reference for analyzing the general solution.

For a general TM, the solution of Eq.~(\ref{ceq3}) reads
\be
\hat{b}(t)=\sum_{j}\int_{-T/2}^{T/2} g_j(t,t')\hat{a}^{in}_j(t')dt',
\ee
with kernel 
\bal
g_j(t,t')&=-[\calA+u(t-t')] h_j(t'){\rm e}^{f(t,t')},
\label{g1m}
\eal
where $\calA=\exp[-(\gamma T+\eta^2)/2]\big/\!\left(1-\exp[-(\gamma T+\eta^2)/2]\right)$, 
$f(t,t')=-\int_{t'}^{t}(\gamma/2+\eta^2|\beta(t'')|^2/2)\,dt''$,
$h_s(t)=\eta\beta(t)$, $h_i(t)=\sqrt{\gamma}$, and $u$ is the unit step function. 
To spectrally decorrelate the signal and idler fields, one seeks to factorize $g_j(t,t')$, which requires $\mathcal{A}\gg1$ and $|f(t,t')|\ll1$. 
These conditions are satisfied in the regime $\eta\sim\sqrt{\gamma T}\to0$ ($\gamma/\Delta\omega\to0$), where $g_j(t,t')=-2h_j(t')/(\gamma T+\eta^2)$. The idler output in frequency domain becomes $\hat{a}_i^{out}(0)=\mu'_0\hat{A}_s^{in}(0)+\nu'_{0}\hat{a}^{in}_{i}(0)$ and $\hat{a}_i^{out}(\omega_{n_i})=\hat{a}_i^{in}(\omega_{n_i})$ for $n_i\neq 0$, where $\mu'_0=-2\eta\sqrt{\gamma T}/(\gamma T+\eta^2)$ and $\nu'_0=(\eta^2-\gamma T)/(\gamma T+\eta^2)$. In particular, when $\eta=\sqrt{\gamma T}$, Eq.~(\ref{fullc}) is recovered. Such factorization is impossible when the field is treated in the $T \to \infty$ limit (i.e.\ under a continuous-frequency expansion), since $\sqrt{\gamma T}$ inevitably diverges. In contrast, for finite $T$, where the spectrum discretizes into bins of width $\Delta\omega = 2\pi/T$, a sufficiently narrow cavity bandwidth ($\gamma \ll \Delta\omega$) can isolate a single mode for conversion. Fig.~\ref{fig:4}(a) shows the amplitude of the frequency-domain transfer function $G_s(\omega_{n_i},\omega_{n_s})$, proportional to the Fourier transform of $g_s(t,t')$ \cite{supp}, for $\gamma/\Delta\omega=\eta^{2}/2\pi=10^{-2}$. Using the same parameter settings as in (a) but varying $\gamma/\Delta\omega$ under the constraint $\eta=\sqrt{\gamma T}$, Fig.~\ref{fig:4}(b) shows the separability of the target TM and its fidelity to $\beta(t)$, both exceeding $0.995$ for $\gamma/\Delta\omega\le 0.1$. Setting $\gamma/\Delta\omega=0.1$, the pump bandwidth constraint $\Omega_p<\Omega_{FSR}$ limits the number of frequency bins in the target TM to $N=\Omega_p/\Delta\omega< F/10$, where $F=\Omega_{ FSR}/\gamma$ is the cavity finesse. For state-of-the-art values $F\sim10^6$, the target TM can span up to $N\sim10^5$ frequency components, enabling the high spectral resolution required for applications such as quantum radar \cite{Chen:25,PhysRevApplied.21.034004}.

To account for internal losses, we model coupling to an ancillary bath mode $\hat{d}^{in}$ with loss rate $\iota$. For a general TM, taking the limit $\eta\sim\sqrt{(\gamma+\iota) T}\to 0$, the idler output in the frequency domain is found to be
$\hat{a}_i^{out}(0)=\mu''_0\hat{A}_s^{in}(0)+\nu''_{0}\hat{a}^{in}_{i}(0)+\upsilon''_{0}\hat{d}^{in}(0)$ and $\hat{a}_i^{out}(\omega_{n_i})=\hat{a}_i^{in}(\omega_{n_i})$ for $n_i\neq 0$, with coefficients $\mu''_0=-2\eta\sqrt{\gamma T}/(\gamma T+\iota T+\eta^2)$, $\nu''_0=(\eta^2-\gamma T+\iota T)/(\gamma T+\iota T+\eta^2)$ and $\upsilon''_0=-2\sqrt{\gamma\iota}/(\gamma T+\iota T+\eta^2)$ \cite{supp}. 
The CE, $|\mu_0''|^2$, of the target TM reaches its maximum value of $1/(1+\iota/\gamma)$ at $\eta = \sqrt{(\gamma+\iota)T}$, as shown in Fig.~\ref{fig:4}(b), highlighting that internal losses lower the achievable peak CE.

The QPG based on the CSFG can be naturally extended to MQPG. Let the cavity resonances be $\omega_{i,c,m}=\omega_{i,c}+m\Omega_{FSR}$, and denote the corresponding idler modes by $\hat{a}_{i,m}$, each with bandwidth $\Omega_{FSR}$. The signal field $\hat{a}_s$ is assumed to have the same bandwidth. The pump field is proportional to $\beta(t)=\sum_m\beta_m(t){\rm exp}(-i\omega_{p,c,m}t)$, where each tone $\beta_m(t){\rm exp}(-i\omega_{p,c,m}t)$ is narrowband (bandwidth<$\Omega_{FSR}$) and centered at $\omega_{p,c,m}=\omega_{p,c}+m\Omega_{FSR}$. The pump envelopes $\{\beta_m(t)\}$ are mutually orthogonal. The above setup enables a mode-by-mode treatment \cite{supp}. 
In the limit $\eta=\sqrt{\gamma T}\to 0$, the idler output in the frequency domain for each resonance is $\hat{a}_{i,m}^{out}(0)=-\hat{A}_{s,m}^{in}(0)$ and $\hat{a}_{i,m}^{out}(\omega_{n_i})=\hat{a}_{i,m}^{in}(\omega_{n_i})$ for $n_i\neq 0$, with $\hat{A}^{in}_{s,m}(0)=\sum_{n_s}\beta_m(\omega_{-n_s})\hat{a}^{in}_{s}(\omega_{n_s})$. This realizes a multiport frequency-bin interferometer for linear optical quantum networks \cite{PRXQuantum.5.040329} with the idler output $\hat{a}_{i,m}^{out}(0)=\sum_l U_{ml}\hat{a}^{in}_{s}(\omega_{l})$, where $U_{ml}=-\beta_m(\omega_{-l})$. Fig.~\ref{fig:4}(c) shows the amplitude of the three-output transfer function of a three-peak MQPG.

In conclusion, we have demonstrated that optical resonators enable powerful new applications in nonlinear optics, exemplified by cavity-enhanced SPDC for generating spectrally uncorrelated photon pairs and cavity-assisted SFG for quantum pulse gating. Resonators act as versatile mode selectors, allowing efficient and high-resolution control of spectral correlations. Our approach provides a general theoretical framework and practical tools for extending resonator-based techniques to a broad class of nonlinear processes.

\appendix






\maketitle

\section{Review of the interaction and Heisenberg picture}
\label{repicture}
The Hamiltonian in Schrödinger picture is given by
\be
\hat{H}_S(t)=\hat{H}_{0,S}+\hat{H}_{1,S}(t).
\ee
Let $\ket{\psi_S(t)}=U_S(t,t_0)\ket{\psi_S(t_0)}$ be the time-dependent state vector in the Schrödinger picture, where $U_S(t,t_0)=\calT{\rm e}^{-i\int^t_{t_0}\hat{H}_{S}(t')dt'/\hbar}$. 
The state in the interaction picture is 
\bal
\ket{\psi_I(t)}&=U^{\dagger}_{0,S}(t,t_0)\ket{\psi_S(t)}\\&=U^{\dagger}_{0,S}(t,t_0)U_S(t,t_0)\ket{\psi_S(t_0)},
\label{psii1}
\eal
where $U_{0,S}(t,t_0)=\calT{\rm e}^{-i\int^t_{t_0}\hat{H}_{0,S}(t')dt'/\hbar}$.
 An operator in the interaction picture is defined as
\be
\hat{A}_I(t)=U^{\dagger}_{0,S}(t,t_0)\hat{A}_S(t)U_{0,S}(t,t_0),
\ee
where $\hat{A}_S(t)$ will typically not depend explicitly on $t$ and can be rewritten as just $\hat{A}_S$. Alternatively, $\hat{A}_I(t)$ can be obtained by solving
\be
\dot{\hat{A}}_I(t)=U^{\dagger}_{0,S}(t,t_0)\biggl\{\frac{i}{\hbar}[\hat{H}_{0,S}(t),\hat{A}_S(t)]+\frac{\partial \hat{A}_S(t)}{\partial t}\biggr\} U_{0,S}(t,t_0).
\ee
If $\hat{A}_S$ has no explicit dependence on time, the term $\partial \hat{A}_S(t)/\partial t=0$. The time evolution of the state in the interaction picture is governed by
\be
i\hbar\frac{d}{dt}\ket{\psi_I(t)}=\hat{H}_{1,I}(t)\ket{\psi_I(t)},
\ee
where $\hat{H}_{1,I}(t)=U^{\dagger}_{0,S}(t,t_0)\hat{H}_{1,S}(t)U_{0,S}(t,t_0)$. The solution is \bal
\ket{\psi_I(t)}&=U_{1,I}(t,t_0)\ket{\psi_I(t_0)}=U_{1,I}(t,t_0)\ket{\psi_S(t_0)},
\label{psii2}
\eal
where $U_{1,I}(t,t_0)=\calT{\rm e}^{-i\int^t_{t_0}\hat{H}_{1,I}(t')dt'/\hbar}$. Comparing Eq. (\ref{psii1}) and (\ref{psii2}) gives $U_S(t,t_0)=U_{0,S}(t,t_0)U_{1,I}(t,t_0)$, so that the relation between an operator in the interaction and Heisenberg pictures is
\be
\hat{A}_H(t)=U^{\dagger}_{1,I}(t,t_0)\hat{A}_I(t)U_{1,I}(t,t_0),
\label{hir}
\ee
where $\hat{A}_H(t)=U^{\dagger}_S(t,t_0)\hat{A}_S(t)U_S(t,t_0)$ is the operator in the Heisenberg picture, satisfying the Heisenberg equation of motion
\be
\dot{\hat{A}}_H(t)=U^{\dagger}_{S}(t,t_0)\biggl\{\frac{i}{\hbar}[\hat{H}_{S}(t),\hat{A}_S(t)]+\frac{\partial \hat{A}_S(t)}{\partial t}\biggr\} U_{S}(t,t_0).
\ee

\section{Detailed analysis of the CSPDC}
\label{CSPDC}
In this section, we present a detailed analysis of the CSPDC results discussed in the main text.
The total Hamiltonian of CSPDC reads
\bal
\hat{H}/\hbar&=\sum_j\big[\omega_{j,c}\hat{b}^{\dagger}_j\hat{b}_j+\int d\omega_j (\omega_j+\omega_{j,c})\hat{a}_j^{\dagger}(\omega_j)\hat{a}_j(\omega_j)+i\sqrt{\gamma_j} (\hat{a}_j^{\dagger}\hat{b}_j-\hat{b}_j^{\dagger}\hat{a}_j)\big]-i \eta [\hat{b}_s^{\dagger}\hat{b}_i^{\dagger}\beta(t)-\hat{b}_s\hat{b}_i\beta^{*}(t)].
\label{Hamitonianall}
\eal
Here, $j \in \{s,i\}$ labels the signal ($s$) and idler ($i$) modes. The operators $\hat{b}_j$ represent the intracavity modes with central frequencies $\omega_{j,c}$, while $\hat{a}_j$ denote the external field modes coupled to the cavities. The function $\beta(t)$ denotes the scaled spectral profile of the pump field, specified in the frequency domain as
\be
\beta(\omega_p)=\Pi(\omega_p/\Omega_p)=
\begin{cases}
      1, & \text{if}\ |\omega_p|\leq \Omega_p/2 \\
      0, & \text{if}\ |\omega_p|>\Omega_p/2,
    \end{cases}
\ee
where $\Omega_p \gg \max(\gamma_s,\gamma_i)$ is the pump bandwidth. Transforming to the rotating frame, with $\hat{o}_{j}(\omega_j)\to \hat{o}_{j}(\omega_j)e^{-i\omega_{j,c} t}$ and $\beta(t)\to \beta(t)e^{-i\omega_{p,c} t}$ ($o\in\{a,b\}$, $\omega_{p,c}=\omega_{s,c}+\omega_{i,c}$), the Hamiltonian becomes
\bal
\hat{H}=\hat{H}_0+\hat{H}_1,
\label{Hamitonianallro}
\eal
where $\hat{H}_0/\hbar=\sum_{j}[\int d\omega_j \omega_j\hat{a}^{\dagger}_{j}(\omega_j)\hat{a}_{j}(\omega_j)+i\sqrt{\gamma_j} (\hat{a}_j^{\dagger}\hat{b}_j-\hat{b}_j^{\dagger}\hat{a}_j)]$ and
$\hat{H}_1/\hbar=-i \eta [\hat{b}_s^{\dagger}\hat{b}_i^{\dagger}\beta(t)-\hat{b}_s\hat{b}_i\beta^{*}(t)]$. For conciseness, Schr\"{o}dinger-picture subscripts $S$ are omitted. From the Hamiltonian $\hat{H}_0$, one derives the equations of motion for the cavity modes $\hat{b}_j$ in the interaction picture \cite{PhysRevA.31.3761}.
\be
\dot{\hat{b}}_{j,I}(t)=-\frac{\gamma_j}{2}\hat{b}_{j,I}(t)-\sqrt{\gamma_j}\hat{a}_{j,I}^{in}(t),
\label{ceq22}
\ee
where $\hat{a}_{j,I}^{in}(t)=(1/\sqrt{2\pi})\int d\omega_j\hat{a}_{j,I}^{in}(\omega) {\rm e}^{-i\omega t}$. For convenience, we recall the definitions of the input and output modes \(\hat{a}_{j,q}^{in(out)}(\omega_j) = \lim_{t \to \mp \infty} \hat{a}_{j,q}(\omega_j,t) e^{i \omega_j t}\) 
(\(q \in \{I,H\}\)), and \(\hat{a}_{j,S}^{in}(\omega_j) = \lim_{t \to -\infty} \hat{a}_{j,S}(\omega_j) e^{i \omega_j t}\), which satisfy
\(\hat{a}_{j,S}^{in}(\omega_j) = \hat{a}_{j,I}^{in}(\omega_j) = \hat{a}_{j,H}^{in}(\omega_j)\). By solving Eq.~(\ref{ceq22}) in the frequency domain, we obtain
\be
\hat{b}_{j,I}(\omega_j)=\sqrt{2\pi}\xi^-_{j}(\omega_j)\hat{a}^{in}_{j,I}(\omega_j),
\label{cavityra}
\ee
together with the input–output relation linking the intracavity field to the external modes,
\be
\hat{a}^{out}_{j,I}(t)-\hat{a}^{in}_{j,I}(t)=\sqrt{\gamma_j}\hat{b}_{j,I}(t),
\label{inou8}
\ee
we obtain
\be
\hat{a}^{out}_{j,I}(\omega_j)=\frac{\xi^{-}_{j}(\omega_j)}{\xi^{+}_{j}(\omega_j)}\hat{a}^{in}_{j,I}(\omega_j),
\label{inoutr2a}
\ee
where $\xi^{\mp}_{j}(\omega_j)=\sqrt{\gamma_j/2\pi}/(i\omega_j\mp\gamma_j/2)$. In the weak-perturbation limit, the time-evolution operator in the interaction picture is
\bal
U_{1,I}(\infty,-\infty)&\approx {\rm exp}\bigg(-\frac{i}{\hbar}\int_{-\infty}^{\infty}\hat{H}_{1,I} dt\bigg)\\&={\rm exp}\biggl[\int_{-\infty}^{\infty} \eta \hat{b}_{s,I}(t)\hat{b}_{i,I}(t)\beta^{*}(t) dt-h.c.\biggr] \\&={\rm exp}\biggl[\int_{-\infty}^{\infty} \frac{\eta}{(2\pi)^\frac{3}{2}} \hat{b}_{s,I}(\omega_s)\hat{b}_{i,I}(\omega_i)\Pi(\frac{\omega_p}{\Omega_p}){\rm e}^{-i(\omega_s+\omega_i-\omega_p)t}d\omega_sd\omega_id\omega_p dt-h.c.\biggr]\\&={\rm exp}\biggl[\int_{-\infty}^{\infty} \frac{\eta}{(2\pi)^\frac{1}{2}} \hat{b}_{s,I}(\omega_s)\hat{b}_{i,I}(\omega_i)\Pi(\frac{\omega_p}{\Omega_p})\delta(\omega_s+\omega_i-\omega_p)d\omega_sd\omega_id\omega_p-h.c.\biggr]\\&={\rm exp}\biggl[\int_{-\infty}^{\infty} \frac{\eta}{(2\pi)^\frac{1}{2}} \hat{b}_{s,I}(\omega_s)\hat{b}_{i,I}(\omega_i)\Pi(\frac{\omega_s+\omega_i}{\Omega_p})d\omega_sd\omega_i-h.c.\biggr]\\&\approx{\rm exp}\biggl[\int_{-\infty}^{\infty} \sqrt{2\pi}\eta\xi^-_{s}(\omega_s)\xi^-_{i}(\omega_i)  \hat{a}^{in}_{s,I}(\omega_s)\hat{a}^{in}_{i,I}(\omega_i)d\omega_s d\omega_i-h.c.\biggr]\\&={\rm exp}(\sqrt{2\pi}\eta  \hat{A}_{s,I}^{in}\hat{A}^{in}_{i,I}-h.c.).
\label{uapp}
\eal
In the penultimate step of the derivation, we used Eq. (\ref{cavityra}).
The TMs are defined as $\hat{A}_{j,q'}^{in(out)}=\int \xi^{\mp}_{j}(\omega_j) \hat{a}_{j,q'}^{in(out)}(\omega_j)d\omega_j$ with $q'\in \{S,I,H\}$. Using Eq.~(\ref{inoutr2a}) and the identity $\hat{a}_{j,I}^{in}(\omega_j) = \hat{a}_{j,H}^{in}(\omega_j)= \hat{a}_{j,S}^{in}(\omega_j)$, we obtain $\hat{A}_{j,I}^{out}=\hat{A}_{j,I}^{in}=\hat{A}_{j,H}^{in}=\hat{A}_{j,S}^{in}$. According to Eq.~(\ref{hir}), the corresponding Heisenberg-picture mode operator evolves as 
\bal
\hat{A}_{j,H}^{out}&=U_{1,I}^\dagger(\infty,-\infty)\hat{A}_{j,I}^{out}U_{1,I}(\infty,-\infty)={\rm cosh}(\sqrt{2\pi}\eta)\hat{A}^{in}_{j,H}-{\rm sinh}(\sqrt{2\pi}\eta)\hat{A}^{in\dagger}_{\Bar{j},H},
\label{inoutra}
\eal
where $\bar{j}$ denotes the complementary mode ($\bar{s}=i$ and $\bar{i}=s$). From Eq.(\ref{inoutra}), the Schr\"{o}dinger-picture evolution operator is $U_S(\infty,-\infty)=\exp(\sqrt{2\pi}\eta  \hat{A}^{in}_{s,S}\hat{A}^{in}_{i,S}-h.c.)$. The corresponding output state is $\ket{\Phi}\approx\sqrt{1-2\pi\eta^2}\ket{vac}-\sqrt{2\pi}\eta\ket{1_s,1_i}$.
\subsection{Effect of internal loss}
We account for internal losses by introducing ancillary bath modes $\hat{d}_j^{in}$ with loss rate $\iota_j$, where $\iota_j$ is taken to be smaller than the cavity decay rate $\gamma_j$. Eq. (\ref{ceq22}) then becomes
\be
\dot{\hat{b}}_{j,I}(t)=-\frac{\gamma_j+\iota_j}{2}\hat{b}_{j,I}(t)-\sqrt{\gamma_j}\hat{a}_{j,I}^{in}(t)-\sqrt{\iota_j}\hat{d}_{j,I}^{in}(t),
\label{ceq4}
\ee
which, in the frequency domain, is solved as
\bal
\hat{b}_{j,I}(\omega_j)&=\tau_j(\omega_j)\hat{a}_{j,I}^{in}(\omega_j)+\sigma_j(\omega_j)\hat{d}_{j,I}^{in}(\omega_j),
\eal
with $\tau_j(\omega_j)=\sqrt{\gamma_j}/(i\omega_j-\gamma_j/2-\iota_j/2)$ and $\sigma_j(\omega_j)=\sqrt{\iota_j}/(i\omega_j-\gamma_j/2-\iota_j/2)$. Combining with the input–output relation Eq.~(\ref{inou8}), gives
\bal
\hat{a}^{out}_{j,I}(\omega_j)&=t_j(\omega_j)\hat{a}_{j,I}^{in}(\omega_j)+r_j(\omega_j)\hat{d}_{j,I}^{in}(\omega_j),
\label{anoi}
\eal
where $t_j(\omega_j)=(i\omega_j+\gamma_j/2-\iota_j/2)/(i\omega_j-\gamma_j/2-\iota_j/2)$ and $r_j(\omega_j)=\sqrt{\gamma_j\iota_j}/(i\omega_j-\gamma_j/2-\iota_j/2)$. Applying the same method used in solving Eq.~(\ref{uapp}), the time-evolution operator is $U_{1,I}(\infty,-\infty)\approx{\rm exp}(S)$ where
\bal
S=\frac{\eta}{\sqrt{2\pi}} \hat{b}_{s,I}(\omega_s)\hat{b}_{i,I}(\omega_i)-h.c..
\eal
As dictated by Eq.~(\ref{hir}), the corresponding Heisenberg-picture operator is then
\bal
\hat{a}^{out}_{j,H}(\omega_j)&=U_{1,I}^\dagger(\infty,-\infty)\hat{a}_{j,I}^{out}(\omega_j)U_{1,I}(\infty,-\infty)\\&\approx \hat{a}_{j,I}^{out}(\omega_j)+[\hat{a}_{j,I}^{out}(\omega_j),S]\\&=t_j(\omega_j)\hat{a}_{j,I}^{in}(\omega_j)+r_j(\omega_j)\hat{d}_{j,I}^{in}(\omega_j)+\eta\tilde{\xi}^-_j(\omega_j)[\tau_{\bar{j}}^*(\omega_{\bar{j}})\hat{a}_{\bar{j},I}^{in\dagger}(\omega_{\bar{j}})+\sigma_{\bar{j}}^*(\omega_{\bar{j}})\hat{d}_{\bar{j},I}^{in\dagger}(\omega_{\bar{j}})],
\label{Anoia}
\eal
with
\bal
\tilde{\xi}^-_j(\omega_j)&=\frac{t_j(\omega_j)\tau^*_j(\omega_j)+r_j(\omega_j)\sigma^*_j(\omega_j)}{\sqrt{2\pi}}=\frac{\sqrt{\gamma_j/2\pi}}{i\omega_j-\gamma_j/2-\iota_j/2}.
\eal
Hence, the biphoton wavefunction (JSF) in the Heisenberg picture is
\bal
\calJ(\omega_s,\omega_i)&=\bra{0}\hat{a}^{out}_{s,H}(\omega_s)\hat{a}^{out}_{i,H}(\omega_i)\ket{0}\approx-\sqrt{2\pi}\eta\tilde{\xi}^-_s(\omega_s)\tilde{\xi}^-_i(\omega_i),
\eal
and the corresponding biphoton rate is
\bal
p_{si}&=\int |\calJ(\omega_s,\omega_i)|^2d\omega_sd\omega_i=2\pi\eta^2\frac{\gamma_s}{\gamma_s+\iota_s}\frac{\gamma_i}{\gamma_i+\iota_i}.
\eal
The single-photon rate is
\bal
p_j&=\int \bra{0}\hat{a}^{out\dagger}_{j,H}(\omega_j)\hat{a}^{out}_{j,H}(\omega_j)\ket{0}d\omega_sd\omega_i=2\pi\eta^2\frac{\gamma_j}{\gamma_j+\iota_j},
\eal
which gives the total probability of detecting a photon in channel $j$, independent of whether its partner in $\bar{j}$ is present. Therefore, the heralding rate is $\calR_j=p_{si}/p_j={\gamma_{\bar{j}}}/(\gamma_{\bar{j}}+\iota_{\bar{j}})$.

\subsection{Single-cavity configuration}
We consider a variant of the CSPDC scheme in which only the signal channel is cavity-enhanced, and present a detailed analysis. The total Hamiltonian is
\bal
\hat{H}/\hbar&=\omega_{s,c}\hat{b}^{\dagger}_s\hat{b}_s+\sum_j\big[\int d\omega_j (\omega_j+\omega_{j,c})\hat{a}_j^{\dagger}(\omega_j)\hat{a}_j(\omega_j)\big]+i\sqrt{\gamma_s} (\hat{a}_s^{\dagger}\hat{b}_s-\hat{b}_s^{\dagger}\hat{a}_s)-i \eta [\hat{b}_s^{\dagger}\hat{a}_i^{\dagger}\beta(t)-\hat{b}_s\hat{a}_i\beta^{*}(t)].
\label{Hamitonianall}
\eal
The idler mode $\hat{a}_i$ is not cavity-enhanced but directly participates in the nonlinear interaction. $\eta>0$ is the effective nonlinear coupling strength. In the rotating frame defined by the transformations $\hat{o}_{j}(\omega_j)\rightarrow\hat{o}_{j}(\omega_j) {\rm e}^{-i\omega_{j,c} t}$, $\beta(t)\rightarrow\beta(t){\rm e}^{-i\omega_{p,c} t}$  with $o\in\{a,b\}$ and $\omega_{p,c} = \omega_{s,c} + \omega_{i,c}$, the Hamiltonian simplifies to
\bal
\hat{H}=\hat{H}_0+\hat{H}_1,
\label{Hamitonianallro}
\eal
where 
$\hat{H}_0/\hbar=\sum_{j}[\int d\omega_j \omega_j\hat{a}^{\dagger}_{j}(\omega_j)\hat{a}_{j}(\omega_j)]+i\sqrt{\gamma_s} (\hat{a}_s^{\dagger}\hat{b}_s-\hat{b}_s^{\dagger}\hat{a}_s)$
and $\hat{H}_1/\hbar=-i \eta [\hat{b}_s^{\dagger}\hat{a}_i^{\dagger}\beta(t)-\hat{b}_s\hat{a}_i\beta^{*}(t)]$.
From $\hat{H}_0$, the equations of motion for the interaction-picture modes $\hat{a}_{i}(\omega_i)$ and $\hat{b}_s$ follow as \cite{PhysRevA.31.3761}
\be
\dot{\hat{b}}_{s,I}(t)=-\frac{\gamma_s}{2}\hat{b}_{s,I}(t)-\sqrt{\gamma_s}\hat{a}_{s,I}^{in}(t),
\label{ceq1c22}
\ee
\be
\dot{\hat{a}}_{i,I}(\omega_i,t)=-i\omega_i\hat{a}_{i,I}(\omega_i,t).
\label{a1c}
\ee
Solving Eq.~(\ref{ceq1c22}) in the frequency domain gives
\be
\hat{b}_{s,I}(\omega_s)=\sqrt{2\pi}\xi^-_{s}(\omega_s)\hat{a}^{in}_{s,I}(\omega_s),
\label{cavityrsa}
\ee
which, combined with the input–output relation,
\be
\hat{a}^{out}_{s,I}(t)-\hat{a}^{in}_{s,I}(t)=\sqrt{\gamma_s}\hat{b}_{s,I}(t),
\label{inou39}
\ee
yields
\be
\hat{a}^{out}_{s,I}(\omega_s)=\frac{\xi^{-}_{s}(\omega_s)}{\xi^{+}_{s}(\omega_s)}\hat{a}^{in}_{s,I}(\omega_s),
\label{inoutr2sa}
\ee
From Eq.~(\ref{a1c}), the idler evolves as
\be
\hat{a}_{i,I}(\omega_i,t)=\hat{a}^{in}_{i,I}(\omega_i){\rm e}^{-i\omega_i t}=\hat{a}^{out}_{i,I}(\omega_i){\rm e}^{-i\omega_i t},
\ee
implying
\be
\hat{a}^{in}_{i,I}(\omega_i)=\hat{a}^{out}_{i,I}(\omega_i).
\label{inouti1c}
\ee
The time-domain operator is therefore
\be
\hat{a}_{i,I}(t)=(1/\sqrt{2\pi})\int d\omega_i \hat{a}_{i,I}(\omega_i,t)=(1/\sqrt{2\pi})\int d\omega_i \hat{a}^{in(out)}_{i,I}(\omega_i){\rm e}^{-i\omega_i t}.
\ee
The time-evolution operator in the interaction picture is then approximated as
\bal
U_{1,I}(\infty,-\infty)&\approx {\rm exp}\bigg(-\frac{i}{\hbar}\int_{-\infty}^{\infty}\hat{H}_{1,I} dt\bigg)\\&={\rm exp}\biggl[\int_{-\infty}^{\infty} \eta \hat{b}_{s,I}(t)\hat{a}_{i,I}(t)\beta^{*}(t) dt-h.c.\biggr] \\&={\rm exp}\biggl[\int_{-\infty}^{\infty} \frac{\eta}{(2\pi)^\frac{3}{2}} \hat{b}_{s,I}(\omega_s)\hat{a}^{in}_{i,I}(\omega_i)\beta(\omega_p){\rm e}^{-i(\omega_s+\omega_i-\omega_p)t}d\omega_sd\omega_id\omega_p dt-h.c.\biggr]\\&={\rm exp}\biggl[\int_{-\infty}^{\infty} \frac{\eta}{(2\pi)^\frac{1}{2}} \hat{b}_{s,I}(\omega_s)\hat{a}^{in}_{i,I}(\omega_i)\beta(\omega_s+\omega_i)d\omega_sd\omega_i-h.c.\biggr]\\&\approx{\rm exp}\biggl[\int_{-\infty}^{\infty} \eta\xi^-_{s}(\omega_s)\beta(\omega_i)  \hat{a}^{in}_{s,I}(\omega_s)\hat{a}_{i,I}^{in}(\omega_i)d\omega_s d\omega_i-h.c.\biggr]\\&={\rm exp}(\eta  \hat{A}_{s,I}^{in}\hat{A}^{in}_{i,I}-h.c.).
\eal
The TMs are defined as $\hat{A}_{s,q'}^{in(out)}=\int \xi^{\mp}_{s}(\omega_s) \hat{a}^{in(out)}_{s,q'}(\omega_s)d\omega_s$ and $\hat{A}_{i,q'}^{in(out)}=\int \beta(\omega_i) \hat{a}_{i,q'}^{in(out)}(\omega_i)d\omega_i$.
From Eqs.~(\ref{inoutr2sa}) and (\ref{inouti1c}), we find that $\hat{A}_{j,I}^{out}=\hat{A}_{j,I}^{in}=\hat{A}_{j,H}^{in}=\hat{A}_{j,S}^{in}$. Accordingly, the Heisenberg-picture input–output relation is $\hat{A}_{j,H}^{out}={\rm cosh}(\eta)\hat{A}^{in}_{j,H}+{\rm sinh}(\eta)\hat{A}^{in\dagger}_{\Bar{j},H}$. Thus, the Schr\"{o}dinger-picture evolution operator can be written as $
U_{S}(\infty,-\infty)\approx{\rm exp}(\eta  \hat{A}^{in}_{s,S}\hat{A}^{in}_{i,S}-h.c.)
$.

\subsection{Detailed analysis of the CSFG}
In this section we present a detailed analysis of the cavity-enhanced QPG. We expand all fields over a finite observation window $T$, which discretizes the spectrum into frequency bins with spacing $\Delta\omega=2\pi/T$. The total Hamiltonian reads
\bal
\hat{H}/\hbar&=\omega_{i,c}\hat{b}^{\dagger}\hat{b}+\sum_{j,n_j}  [(\omega_{n_j}+\omega_{j,c})\hat{a}^{\dagger}_j(\omega_{n_j})\hat{a}_j(\omega_{n_j})]+i\sqrt{\gamma} (\hat{a}_i^{\dagger}\hat{b}-\hat{b}^{\dagger}\hat{a}_i)-i \eta [\hat{a}_s\hat{b}^{\dagger}\beta(t)-\hat{a}_s^{\dagger}\hat{b}\beta^{*}(t)].
\label{Hamitonianall}
\eal
Here, $\hat{a}_s$ denotes the signal mode, $\hat{b}$ represents the intracavity mode in the idler channel with central frequency $\omega_{i,c}$, and $\hat{a}_i$ is the external field mode coupled to the cavity. The pump envelope $\beta(t)$ is normalized such that $\int_{-T/2}^{T/2}|\beta(t)|^2dt=1$. The mode operators $\hat{a}_j=\sqrt{1/T}\sum_{n_j} \hat{a}_{j}(\omega_{n_j})$ with $\omega_{n_j}=n_j\Delta\omega$. Transforming to the rotating frame $\hat{a}_{j}(\omega_{n_j})\rightarrow\hat{a}_{j}(\omega_{n_j}){\rm e}^{-i\omega_{j,c} t}$, $\hat{b}\rightarrow\hat{b}{\rm e}^{-i\omega_{i,c} t}$, $\beta(t)\rightarrow\beta(t){\rm e}^{-i\omega_{p,c} t}$ with $\omega_{s,c}+\omega_{p,c}=\omega_{i,c}$, the Hamiltonian simplifies to
\bal
\hat{H}=\hat{H}_0+\hat{H}_1,
\label{Hamitonianallro2a}
\eal
where
$
\hat{H}_0/\hbar=\sum_{j,n_j} \omega_{n_j}\hat{a}^{\dagger}_{j}(\omega_{n_j})\hat{a}_{j}(\omega_{n_j})+i\sqrt{\gamma} (\hat{a}^{\dagger}_i\hat{b}-\hat{b}^{\dagger}\hat{a}_i)
$
and
$
\hat{H}_1/\hbar=-i \eta [\hat{a}_s\hat{b}^{\dagger}\beta(t)-\hat{a}_s^{\dagger}\hat{b}\beta^{*}(t)]
$.
\subsection{Interaction-picture analysis in the low-conversion regime.}
We begin by treating the CSFG process perturbatively in the interaction picture under the low-conversion assumption. From $\hat{H}_0$, the equations of motion for $\hat{a}_j(\omega_{n_j})$ and $\hat{b}$ in the interaction picture follow as
\be
\dot{\hat{a}}_{s,I}(\omega_{n_s},t)=-i\omega_{n_s}\hat{a}_{s,I}(\omega_{n_s},t),
\label{aseq}
\ee
\be 
\dot{\hat{a}}_{i,I}(\omega_{n_i},t)=-i\omega_{n_i}\hat{a}_{i,I}(\omega_{n_i},t)+\sqrt{\frac{\gamma}{T}}\hat{b}_I(t),
\label{ai_sol}
\ee
\be
\dot{\hat{b}}_I(t)=-\sqrt{\gamma}\hat{a}_{i,I}(t).
\label{ceq1}
\ee
Solving Eqs.~\eqref{aseq} and \eqref{ai_sol}, we obtain \cite{PhysRevA.31.3761}
\be
\hat{a}_{s,I}(\omega_{n_s},t)={\rm e}^{-i\omega_{n_s} (t+\frac{T}{2})}\hat{a}_{s,I}(\omega_{n_s},-\frac{T}{2}),
\label{awsa}
\ee
\be 
\hat{a}_{i,I}(\omega_{n_i},t)={\rm e}^{-i\omega_{n_i} (t+\frac{T}{2})}\hat{a}_{i,I}(\omega_{n_i},-\frac{T}{2})+\sqrt{\frac{\gamma}{T}}\int_{-\frac{T}{2}}^t{\rm e}^{-i\omega_{n_i} (t-t')}\hat{b}_I(t')dt',
\label{bt_sol}
\ee
yielding
\bal
\hat{a}_{s,I}(t)&=\frac{1}{\sqrt{T}}\sum_{n_s}\hat{a}_{s,I}(\omega_{n_s},t)=\hat{a}_{s,I}^{in}(t),
\label{amode}
\eal
\bal
\hat{a}_{i,I}(t)&=\frac{1}{\sqrt{T}}\sum_{n_i}\hat{a}_{i,I}(\omega_{n_i},t)=\hat{a}_{i,I}^{in}(t)+\frac{\sqrt{\gamma}}{2}\hat{b}_I(t),
\label{bmode1}
\eal
where $\hat{a}_{j,I}^{in(out)}(t)=(1/\sqrt{T})\sum_{n_j} \hat{a}^{in(out)}_{j,I}(\omega_{n_j}) {\rm e}^{-i\omega_{n_j} t}$, with $\hat{a}^{in(out)}_{j,q}(\omega_{n_j})=\hat{a}_{j,q}(\omega_{n_j},\mp T/2){\rm e}^{\mp i\omega_{n_j} T/2}$ ($q\in\{I,H\}$) as defined in the main text. Equation~\eqref{awsa} can equivalently be written as
\be
\hat{a}_{s,I}(\omega_{n_s},t)={\rm e}^{-i\omega_{n_s} (t-\frac{T}{2})}\hat{a}_{s,I}(\omega_{n_s},\frac{T}{2}),
\label{awsa2}
\ee
implying
\be\hat{a}^{out}_{s,I}(\omega_{n_j})=\hat{a}^{in}_{s,I}(\omega_{n_j}).
\label{asinout}
\ee
Substituting Eq.~\eqref{bmode1} into Eq.~\eqref{ceq1} yields the Langevin equation for the cavity mode
\be
\dot{\hat{b}}_I(t)=-\frac{\gamma}{2}\hat{b}_I(t)-\sqrt{\gamma}\hat{a}_{i,I}^{in}(t).
\label{ceq2}
\ee
Solving Eq. (\ref{ceq2}) in the frequency domain, we obtain
\be
\hat{b}_{I}(\omega_{n_i})=\frac{\sqrt{\gamma}}{i\omega_{n_i}-\frac{\gamma}{2}}\hat{a}^{in}_{i,I}(\omega_{n_i}),
\label{cavityrasfg}
\ee
and, using the input–output relation for the intracavity field,
\be
\hat{a}^{out}_{i,I}(t)-\hat{a}^{in}_{i,I}(t)=\sqrt{\gamma_j}\hat{b}_{I}(t),
\label{inou8sfg}
\ee
we have
\be
\hat{a}^{out}_{i,I}(\omega_{n_i})={\rm e}^{i\theta_{n_i}}\hat{a}^{in}_{i,I}(\omega_{n_i}),
\label{aiinout}
\ee
where ${\rm e}^{i\theta_{n_i}}=(i\omega_{n_i}+\gamma/2)/(i\omega_{n_i}-\gamma/2)$. In the limit of weak perturbation, the interaction-picture evolution operator can be approximated as
\bal
U_{1,I}(T/2,-T/2)&\approx {\rm exp}\bigg(-\frac{i}{\hbar}\int_{-T/2}^{T/2}\hat{H}_{1,I} dt\bigg)\\&={\rm exp}\biggl[\int_{-T/2}^{T/2} \eta \hat{a}_{s,I}^\dagger(t)\hat{b}_I(t)\beta^{*}(t) dt-h.c.\biggr] \\&={\rm exp}\biggl[\int_{-T/2}^{T/2} \frac{\eta}{(T)^\frac{3}{2}}\sum_{n_s,n_i,n_p} \hat{a}_{s,I}^{in\dagger}(\omega_{n_s})\hat{b}_I(\omega_{n_i})\beta^*(\omega_{n_p}){\rm e}^{i(\omega_{n_s}-\omega_{n_i}+\omega_{n_p})t} dt-h.c.\biggr]\\&={\rm exp}\biggl[ \frac{\eta}{(T)^\frac{1}{2}} \sum_{n_s,n_i,n_p}\hat{a}_{s,I}^{in\dagger}(\omega_{n_s})\hat{b}_I(\omega_{n_i})\beta^*(\omega_{n_p})\delta_{n_s-n_i+n_p,0}-h.c.\biggr]\\&={\rm exp}\biggl[ \frac{\eta}{(T)^\frac{1}{2}} \sum_{n_s,n_i}\frac{\sqrt{\gamma}\beta^*(\omega_{n_i-n_s})}{(i\omega_{n_i}-\frac{\gamma}{2})}  \hat{a}_{s,I}^{in\dagger}(\omega_{n_s})\hat{a}_{i,I}^{in}(\omega_{n_i})-h.c.\biggr]\\&\approx{\rm exp}\big[\sum_{n_i=-\infty}^{\infty}\rho_{n_i} \hat{a}^{in}_{i,I}(\omega_{n_i}) \hat{A}^{in\dagger}_{s,I}(n_i)-h.c.\big],
\eal
where $\rho_{n_i}=\eta\sqrt{\gamma/T}/(i\omega_{n_i}-\gamma/2)$ and $\hat{A}^{in}_{s,q}(n_i)=\sum_{n_s}\beta(\omega_{n_i-n_s})\hat{a}^{in}_{s,q}(\omega_{n_s})$. From Eq. (\ref{asinout}) and (\ref{aiinout}), we have $\hat{a}^{out}_{i,I}(\omega_{n_i})={\rm e}^{i\theta_{n_i}}\hat{a}^{in}_{i,I}(\omega_{n_i})={\rm e}^{i\theta_{n_i}}\hat{a}^{in}_{i,H}(\omega_{n_i})$ and $\hat{A}^{out}_{s,I}(n_i)=\hat{A}^{in}_{s,I}(n_i)=\hat{A}^{in}_{s,H}(n_i)$. As dictated by Eq.~(\ref{hir}), the Heisenberg-picture output operator reads
\bal
\hat{a}^{out}_{i,H}(\omega_{n_i}) &=U_{1,H}^\dagger(T/2,-T/2)\hat{a}^{out}_{i,I}(\omega_{n_i}) U_{1,I}(T/2,-T/2)\\&={\rm e}^{i\theta_{n_i}}U_{1,H}^\dagger(T/2,-T/2)\hat{a}^{in}_{i,I}(\omega_{n_i}) U_{1,I}(T/2,-T/2)\\&={\rm cos}(|\rho_{n_i}|){\rm e}^{i\theta_{n_i}}\hat{a}^{in}_{i,I}(\omega_{n_i})+{\rm sin}(|\rho_{n_i}|){\rm e}^{i(\theta_{n_i}+\phi_{n_i})}\hat{A}^{in}_{s,I}(n_i)\\&={\rm cos}(|\rho_{n_i}|){\rm e}^{i\theta_{n_i}}\hat{a}^{in}_{i,H}(\omega_{n_i})+{\rm sin}(|\rho_{n_i}|){\rm e}^{i(\theta_{n_i}+\phi_{n_i})}\hat{A}^{in}_{s,H}(n_i)\\&\to{\rm cos}(|\rho_{n_i}|)\hat{a}^{in}_{i,H}(\omega_{n_i})+{\rm sin}(|\rho_{n_i}|)\hat{A}^{in}_{s,H}(n_i).
\eal
In the final step of the calculation, we absorb the phases ${\rm e}^{i\phi_{n_i}}=\rho_{n_i}/|\rho_{n_i}|$ and ${\rm e}^{i\theta_{n_i}}$ into the operators $\hat{a}^{in}_{i,H}(\omega_{n_i})$ and $\hat{A}^{in}_{s,H}(n_i)$ for notational convenience.

\subsection{Heisenberg-picture analysis in the high-conversion regime.}
While the previous discussion treated the CSFG process perturbatively in the interaction picture, we now consider the Heisenberg-picture evolution, enabling a non-perturbative description suitable for higher conversion regimes. Starting from Eq.~\eqref{Hamitonianallro2a}, the Heisenberg equations of motion for $\hat{a}_j(\omega_{n_j})$ and $\hat{b}$ are obtained as
\be
\dot{\hat{a}}_{s}(\omega_{n_s},t)=-i\omega_{n_s}\hat{a}_{s}(\omega_{n_s},t)+\frac{\eta}{\sqrt{T}}\hat{b}(t)\beta^{*}(t),
\label{aseqh}
\ee
\be 
\dot{\hat{a}}_{i}(\omega_{n_i},t)=-i\omega_{n_i}\hat{a}_{i}(\omega_{n_i},t)+\sqrt{\frac{\gamma}{T}}\hat{b}(t),
\label{ai_solh}
\ee
\be
\dot{\hat{b}}(t)=-\eta\hat{a}_s(t)\beta(t)-\sqrt{\gamma}\hat{a}_{i}(t).
\label{ceq1h}
\ee
Solving Eqs.~\eqref{aseqh}–\eqref{ai_solh}, following the method of Ref.~\cite{PhysRevA.31.3761}, yields
\be
\hat{a}_{s}(\omega_{n_s},t)={\rm e}^{-i\omega_{n_s} (t+\frac{T}{2})}\hat{a}_{s}(\omega_{n_s},-\frac{T}{2})+\frac{\eta}{\sqrt{T}}\int_{-\frac{T}{2}}^t {\rm e}^{-i\omega_n(t-t')}\hat{b}(t')\beta^{*}(t')dt',
\ee
\be 
\hat{a}_{i}(\omega_{n_i},t)={\rm e}^{-i\omega_{n_i} (t+\frac{T}{2})}\hat{a}_{i}(\omega_{n_i},-\frac{T}{2})+\sqrt{\frac{\gamma}{T}}\int_{-\frac{T}{2}}^t{\rm e}^{-i\omega_{n_i} (t-t')}\hat{b}(t')dt',
\ee
yielding
\bal
\hat{a}_{s}(t)&=\frac{1}{\sqrt{T}}\sum_{n_s}\hat{a}_{s}(\omega_{n_s},t)=\hat{a}_{s}^{in}(t)+\frac{\eta}{2}\hat{b}(t)\beta^{*}(t),
\label{amode}
\eal
\bal
\hat{a}_{i}(t)&=\frac{1}{\sqrt{T}}\sum_{n_i}\hat{a}_{i}(\omega_{n_i},t)=\hat{a}_{i}^{in}(t)+\frac{\sqrt{\gamma}}{2}\hat{b}(t).
\label{bmode}
\eal
Substituting Eqs. (\ref{amode}) and (\ref{bmode}) into Eq. (\ref{ceq1h}), we obtain the Langevin equation for the cavity mode
\be
\dot{\hat{b}}(t)=-\frac{\gamma}{2}\hat{b}(t)-\eta\hat{a}_s^{in}(t)\beta(t)-\frac{\eta^2}{2}\hat{b}(t)|\beta(t)|^2-\sqrt{\gamma}\hat{a}_i^{in}(t),
\label{ceq2a}
\ee
with periodic boundary condition $\hat{b}(T/2)=\hat{b}(-T/2)$. 

In Ref.~\cite{Chen:25}, we considered a special case in which the pump spectral amplitudes $\beta(\omega_{-n_p})$ are represented by a large set of independent random numbers drawn from a common complex Gaussian distribution. In the limit $M\to\infty$, this yields $|\beta(t)|^2=1/T$, since ${\rm lim}_{M\to\infty}\sum_{k_p=1}^M\beta(\omega_{k_p-n_p})\beta^*(\omega_{k_p-m_p})=\delta_{m_p,n_p}$. Solving Eq. (\ref{ceq2a}) in frequency domain, we find
\be
\hat{b}(\omega_{n_i})=\frac{\frac{\eta}{\sqrt{T}}}{i\omega_{n_i}-\frac{\gamma}{2}-\frac{\eta^2}{2T}}\hat{A}_s^{in}(n_i)+\frac{\sqrt{\gamma}}{i\omega_{n_i}-\frac{\gamma}{2}-\frac{\eta^2}{2T}}\hat{a}^{in}_{i}(\omega_{n_i}),
\label{cn}
\ee
and, using the input–output relation associated with the cavity mode,
\be
\hat{a}_i^{out}(t)-\hat{a}_i^{in}(t)=\sqrt{\gamma}\hat{b}(t),
\label{inoutcaa}
\ee
we obtain
\be
\hat{a}^{out}_{i}(\omega_{n_i})=\frac{\eta\sqrt{\frac{\gamma}{T}}}{i\omega_{n_i}-\frac{\gamma}{2}-\frac{\eta^2}{2T}}\hat{A}_s^{in}(n_i)+\frac{i\omega_{n_i}+\frac{\gamma}{2}-\frac{\eta^2}{2T}}{i\omega_{n_i}-\frac{\gamma}{2}-\frac{\eta^2}{2T}}\hat{a}^{in}_{i}(\omega_{n_i}).
\label{bouta}
\ee
In the limit $\eta=\sqrt{\gamma T}\rightarrow 0$  ($\gamma/\Delta\omega\rightarrow 0$), the output reduces to 
\be
\hat{a}^{out}_{i}(\omega_{n_i})=
\begin{cases}
    -\hat{A}_s^{in}(0) & \text{if} \ n_j=0\\
    \hat{a}^{in}_{i}(\omega_{n_j}) & \text{if} \ n_j\neq0.
\end{cases}
\label{fullca}
\ee
This limiting case provides a useful benchmark for the general solution.

For the general case of TM, the solution of Eq. (\ref{ceq2a}) reads
\be
\hat{b}(t)=\sum_{j}\int_{-T/2}^{T/2} g_j(t,t')\hat{a}^{in}_j(t')dt',
\label{boutgc}
\ee
with the response functions
\bal
g_j(t,t')&=-\big[\frac{{\rm e}^{-\frac{\gamma T}{2}-\frac{\eta^2}{2}}}{1-{\rm e}^{-\frac{\gamma T}{2}-\frac{\eta^2}{2}}}+u(t-t')\big]
 h_j(t'){\rm e}^{-\int^t_{t'}[\frac{\gamma}{2}+\frac{\eta^2}{2}|\beta(t'')|^2]dt''},
\label{g1a}
\eal
where $h_s(t)=\eta\beta(t)$, $h_i(t)=\sqrt{\gamma}$, and $u$ is the unit step function. Transforming to the frequency domain, the intracavity operator can be expressed as
\be
\hat{b}(\omega_{n_i})=\sum_{m_j}\tilde{g}_j(\omega_{n_i},\omega_{m_j})\hat{a}^{in}_j(\omega_{m_j}),
\ee
with 
\be
\tilde{g}_j(\omega_{n_i},\omega_{m_j})=\frac{1}{T}\int_{-T/2}^{T/2} dt\int_{-T/2}^{T/2} dt'{\rm e}^{i\omega_{n_i} t}g_j(t,t') {\rm e}^{-i\omega_{m_j} t'}.
\ee
Combining this with Eq.~(\ref{inoutcaa}), the idler output in the frequency domain becomes
\be
\hat{a}^{out}_i(\omega_{n_i})=\sum_{m_j}G_j(\omega_{n_i},\omega_{m_j})\hat{a}^{in}_j(\omega_{m_j}),
\ee
where the transfer functions are 
\be
G_s(\omega_{n_i},\omega_{m_s})=\sqrt{\gamma}\tilde{g}_s(\omega_{n_i},\omega_{m_s}),\quad
G_i(\omega_{n_i},\omega_{m_i})=\sqrt{\gamma}\tilde{g}_i(\omega_{n_i},\omega_{m_i})+\delta_{{n_i},{m_i}}.
\ee
Applying singular-value decomposition to $G_j$, we write
\be
G_j(\omega_{n_i},\omega_{m_j})=\sum_k\lambda_{j,k} \psi_k(\omega_{n_i})\varphi_{j,k}(\omega_{m_j}),
\ee
with $\sum_j|\lambda_{j,k}|^2=1$ to preserve unitarity. The CE of the $k$-th Schmidt mode is given by $|\lambda_{s,k}|^2$ and the separability of the target TM is defined as $|\lambda_{s,0}|^2/\sum_{k}|\lambda_{s,k}|^2$ \cite{Reddy:13}. These quantities provide a quantitative measure of how effectively the QPG selects the desired target mode.

In the limit $\eta\sim\sqrt{\gamma T}\rightarrow 0$  ($\gamma/\Delta\omega\rightarrow 0$), Eq. (\ref{g1a}) reduces to
\be
g_j(t,t')=-2 h_j(t')/(\gamma T+\eta^2).
\ee
Consequently, the intracavity field becomes
\bal
\hat{b}(t)&=-\int_{-T/2}^{T/2} \frac{2}{\gamma T+\eta^2}\big[\eta\beta(t')\hat{a}_s^{in}(t')+\sqrt{\gamma}\hat{a}_i^{in}(t')\big]dt'\\&=-\frac{2}{\gamma T+\eta^2}[\eta\hat{A}_s^{in}(0)+\sqrt{\gamma T}\hat{a}^{in}_{i}(0)],
\label{ctsol}
\eal
where $\hat{a}^{in}_{i}(0)=(1/\sqrt{T})\int_{-T/2}^{T/2}\hat{a}^{in}_{i}(t)\,dt$ denotes the zero-frequency mode. In the frequency domain, this reduces to
\be
\hat{b}(\omega_{n_i})=
\begin{cases}
    -\frac{2}{\gamma T+\eta^2}[\eta\sqrt{T}\hat{A}_s^{in}(0)+\sqrt{\gamma }T\hat{a}^{in}_{i}(0)] & \text{if} \ n_i=0\\
    0 & \text{if} \ n_i\neq0,
\end{cases}
\label{cwsol}
\ee
and the corresponding idler input–output relation reads
\bal
\hat{a}_i^{out}(\omega_{n_i})=
\begin{cases}
    -\frac{1}{\gamma T+\eta^2}[2\eta\sqrt{\gamma T}\hat{A}_s^{in}(0)+(\gamma T-\eta^2)\hat{a}^{in}_{i}(0)] & \text{if} \ n_i=0\\
    \hat{a}_i^{in}(\omega_{n_i}) & \text{if} \ n_i\neq0.
\end{cases}
\eal

\subsection{Effect of internal loss}
To account for internal losses, we model coupling to an ancillary bath mode $\hat{d}^{in}$ with loss rate $\iota$. Then Eq. (\ref{ceq2a}) becomes
\bal
\dot{\hat{b}}(t)&=-\frac{\gamma}{2}\hat{b}(t)-\frac{\iota}{2}\hat{b}(t)-\eta\hat{a}_s^{in}(t)\beta(t)-\frac{\eta^2}{2}\hat{b}(t)|\beta(t)|^2-\sqrt{\gamma}\hat{a}_i^{in}(t)-\sqrt{\iota}\hat{d}^{in}(t).
\label{ceqloss}
\eal
Applying the same method used for solving Eq.~(\ref{ceq2a}) in the general TM case, and taking the limit $\eta\sim\sqrt{(\gamma+\iota) T}\to 0$, the idler output in the frequency domain is found to be
\bal
\hat{a}^{out}_{i}(\omega_{n_i})=
\begin{cases}
    \mu''_0\hat{A}_s^{in}(0)+\nu''_{0}\hat{a}^{in}_{i}(0)+\upsilon''_{0}\hat{d}^{in}(0) & \text{if} \ n_j=0\\
    \hat{a}^{in}_{i}(\omega_{n_j}) & \text{if} \ n_j\neq0,
\end{cases}
\eal
with coefficients 
\bal
\mu''_0&=-\frac{2\eta\sqrt{\gamma T}}{\gamma T+\iota T+\eta^2}, 
\\ \nu''_0&=\frac{\eta^2-\gamma T+\iota T}{\gamma T+\iota T+\eta^2},
\\ \upsilon''_0&=-\frac{2\sqrt{\gamma\iota}}{\gamma T+\iota T+\eta^2}.\nonumber
\eal 
The CE, $|\mu_0''|^2$, of the target TM reaches its maximum value $
1/(1+\iota/\gamma)$, when $\eta=\sqrt{(\gamma+\iota)T}$.

\subsection{Multi-output QPG}
The MQPG, based on the CSFG process, is governed by the Hamiltonian
\bal
\hat{H}/\hbar&=\sum_m\omega_{i,c,m}\hat{b}_m^{\dagger}\hat{b}_m+\sum_{n_s}  (\omega_{n_s}+\omega_{s,c})\hat{a}_s^{\dagger}(\omega_{n_s})\hat{a}_s(\omega_{n_s})+\sum_{m,n_i}(\omega_{n_i}+\omega_{i,c,m})\hat{a}^{\dagger}_{i,m}(\omega_{n_i})\hat{a}_{i,m}(\omega_{n_i})\\&+i\sqrt{\gamma} (\sum_m\hat{a}_{i,m}^{\dagger}\sum_{m'}\hat{b}_{m'}-\sum_{m'}\hat{b}_{m'}^{\dagger}\sum_{m}\hat{a}_{i,m})-i \eta [\hat{a}_s\sum_m\hat{b}_m^{\dagger}\sum_{m'}\beta_{m'}(t){\rm e}^{-i\omega_{p,c,m'} t}\\&-\hat{a}_s^{\dagger}\sum_m\hat{b}_m\sum_{m'}\beta_{m'}^{*}(t){\rm e}^{i\omega_{p,c,m'} t}],
\label{Hamitonianall}
\eal
where $\omega_{i,c,m}=\omega_{i,c}+m\Omega_{FSR}$, $\omega_{p,c,m}=\omega_{p,c}+m\Omega_{FSR}$, and $\hat{a}_{i,m}=\sqrt{1/T}\sum_{n_i}  \hat{a}_{i,m}(\omega_{n_i})$ denotes the external idler mode, coupled to the cavity, restricted to a bandwidth $\Omega_{FSR}$ around $\omega_{i,c,m}$, associated with intracavity mode $\hat{b}_m$. 

In the rotating frame defined by $\hat{a}_{s}(\omega_{n_s})\rightarrow\hat{a}_{s}(\omega_{n_s}){\rm e}^{-i\omega_{s,c} t}$, $\hat{a}_{i,m}(\omega_{n_i})\rightarrow\hat{a}_{i,m}(\omega_{n_i}){\rm e}^{-i\omega_{i,c,m} t}$ and $\hat{b}_m\rightarrow\hat{b}_m{\rm e}^{-i\omega_{i,c,m} t}$, the Langevin equations for the cavity modes become \cite{PhysRevA.43.543}
\be
\dot{\hat{b}}_k(t)=-\frac{\gamma}{2}\hat{b}_k(t)-\eta\hat{a}_s^{in}(t)\beta_k(t)-\frac{\eta^2}{2}\sum_m\hat{b}_m(t)\beta_m^*(t)\beta_k(t)-\sqrt{\gamma}\hat{a}_{i,k}^{in}(t),
\label{ceqmul}
\ee
where rapidly rotating terms $\propto{\rm exp}(-im\Omega_{FSR}t)$ with $m\neq0$ have been neglected. 

Solving Eq. (\ref{ceqmul}) in the limit $\eta\sim\sqrt{\gamma T}\rightarrow 0$  ($\gamma/\Delta\omega\rightarrow 0$), we obtain an analogous solution to Eq.~(\ref{ctsol}):
\bal
\hat{b}_k(t)=-\frac{2}{\gamma T+\eta^2}[\eta\hat{A}_{s,k}^{in}(0)+\sqrt{\gamma T}\hat{a}^{in}_{i,k}(0)]+K_k,
\label{ctsolm}
\eal
where $\hat{A}^{in}_{s,k}(0)=\int_{-T/2}^{T/2}dt\beta_k(t)\hat{a}_s^{in}(t)=\sum_{n_s}\beta_k(\omega_{-n_s})\hat{a}^{in}_{s}(\omega_{n_s})$, and the additional term 
$K_k\propto\int_{-T/2}^{T/2}dt\sum_{m\neq k}\hat{b}_m(t)\beta_m^*(t)\beta_k(t)
$. Eq.~(\ref{ctsolm}) is time-independent, so only the zero-frequency component survives: $\hat{b}_{k}(\omega_{n_i})=0$ for $n_i\neq 0$ as in Eq.~(\ref{cwsol}).  
Consequently, $\hat{b}_{m}(t)=(1/\sqrt{T})\sum_{n_i}\hat{b}_{m}(\omega_{n_i}){\rm exp}(-i\omega_{n_i}t)=(1/\sqrt{T})\hat{b}_{m}(\omega_{0})$ and the additional term vanishes due to the orthogonality of $\{\beta_m(t)\}$:
\be
K_k\propto\int_{-T/2}^{T/2}dt\sum_{m\neq k}\hat{b}_m(\omega_0)\beta_m^*(t)\beta_k(t)=\sum_{m\neq k}\hat{b}_m(\omega_0)\delta_{m,k}=0.
\ee
Thus, the idler input–output relation for each cavity resonance is
\be
\hat{a}_{i,k}^{out}(\omega_{n_i})=
\begin{cases}
    -\frac{1}{\gamma T+\eta^2}[2\eta\sqrt{\gamma T}\hat{A}_{s,k}^{in}(0)+(\gamma T-\eta^2)\hat{a}^{in}_{i,k}(0)] & \text{if} \ n=0\\
    \hat{a}_{i,k}^{in}(\omega_{n_i}) & \text{if} \ n\neq0.
\end{cases}
\ee
In the special case $\eta=\sqrt{\gamma T}$, the output modes simplify to
\be
\hat{a}^{out}_{i,k}(\omega_{n_i})=
\begin{cases}
    -\hat{A}_{s,k}^{in}(0) & \text{if} \ n_j=0\\
    \hat{a}^{in}_{i,k}(\omega_{n_j}) & \text{if} \ n_j\neq0.
\end{cases}
\ee
This implements a multiport frequency-bin interferometer for linear optical quantum networks \cite{PRXQuantum.5.040329}, with the idler output given by $\hat{a}_{i,m}^{out}(0)=\sum_l U_{ml}\hat{a}^{in}_{s}(\omega_{l})$, where $U_{ml}=-\beta_m(\omega_{-l})$.

\bibliography{apssamp.bib}

\end{document}